\documentclass[letterpaper,10pt,conference]{ieeeconf}
\IEEEoverridecommandlockouts                              
\usepackage{common}
\usepackage{listings}
\usepackage{lstlang1}
\usepackage{lstlang2}
\usepackage{lstlang3}
\usepackage{graphics}
\usepackage{amsmath}
\usepackage{hyperref}
\usepackage{tikz}

\newcommand\copyrighttext{%
  \footnotesize \copyright 2021 IEEE.  Personal use of this material is permitted.  Permission from IEEE must be obtained for all other uses, in any current or future media, including reprinting/republishing this material for advertising or promotional purposes, creating new collective works, for resale or redistribution to servers or lists, or reuse of any copyrighted component of this work in other works.}
\newcommand\copyrightnotice{%
\begin{tikzpicture}[remember picture,overlay]
\node[anchor=south,yshift=10pt] at (current page.south) {\fbox{\parbox{\dimexpr\textwidth-\fboxsep-\fboxrule\relax}{\copyrighttext}}};
\end{tikzpicture}%
}

\title{\LARGE\bf The Resh Programming Language for Multirobot Orchestration}

\newcommand{\opt}{{$O$}}
\newcommand{\supdagger}{$^\dagger$}

\author{Martin Carroll\supdagger\ and Kedar S. Namjoshi\supdagger\ and Itai Segall\supdagger%
\thanks{\supdagger Nokia Bell Labs, Murray Hill, N.J.
  {\tt\small <first>.<last>@ nokia-bell-labs.com}}%
}

\newcommand{\phresh}{\includegraphics[width=1.5ex]{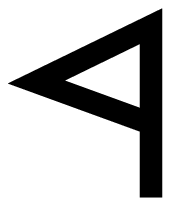}}

\begin{document}

\maketitle
\thispagestyle{plain}
\pagestyle{plain}

\copyrightnotice

\begin{abstract}
This paper describes Resh, a new, statically typed, interpreted programming language and associated runtime for orchestrating multirobot systems.
The main features of Resh are:
(1) It offloads much of the tedious work of programming such systems away from the programmer and into the language runtime;
(2) It is based on a small set of temporal and locational operators; and
(3) It is not restricted to specific robot types or tasks.
The Resh runtime consists of three engines 
that collaborate to run a Resh program using the available robots in their current environment.
This paper describes both Resh and its runtime and gives examples of its use.
\end{abstract}

\section{INTRODUCTION}
\label{sec:intro}

Programming dynamic, heterogeneous teams of robots to perform coordinated tasks is hard.
Even simple tasks can require a daunting amount of logic.
The program must include logic to handle the wide variety of scenarios that can occur while the program is executing.
For example, the program must deal with the following questions:
What robots are currently available for use?
What are their capabilities?
Where is each robot located?
What is the geometry of their environment?
Are they currently engaged in other tasks?
Which robots can perform which parts of which tasks at the same time?
What should be done when a robot is added to or deleted from the pool,
or when a robot completes, or fails to complete, some part of a task?
And so on.
The essential logic of the program is soon overwhelmed by, and intertwined with, code to handle these matters.

Our goal is: To enable the programmer to cleanly express the essential task-sequencing logic, and to take care of all other matters for the programmer. 

This separation can be achieved by burying the nonessential code underneath an API in a traditional programming language.
We take a different approach by introducing Resh,\footnote{Resh is the twentieth letter of the Semitic abjads, the corresponding Phoenician letter \phresh\ of which is a precursor to the modern letter R, which is the first letter of ``robot orchestration."} a domain-specific language for the problem.
The advantage of this approach is that one is free to invent convenient language constructs that may be awkward to emulate with an API\@.
In Resh, the programmer expresses only the task-sequencing logic; other matters are handed by the associated runtime.

The main contributions of this work are:
(1) Resh, a new programming language for orchestrating multirobot systems, based on a small set of temporal and locational operators.
Resh is designed to orchestrate multiple concurrently active, largely asynchronous tasks (i.e., not tightly controlled swarm behavior).
The language is extensible in the types of robots that are supported.
It is deliberately not Turing-complete, to facilitate the automated synthesis of control strategies.
(2) A \emph{geometry-aware} runtime that optimizes the execution of Resh programs.
The runtime takes care of all other logic for the programmer.

The Resh system is in regular use to program a complex and frequently changing multirobot system.
In our experience, Resh has simplified the programming task 
compared with lower-level approaches.



\section{THE RESH LANGUAGE}
\label{sec:resh}

We illustrate the core Resh constructs through small examples that also give a feel for what it is like to program in Resh.
Readers are invited to visit the Resh playground,\footnote{\tt https://resh.io/play}
where programs can be run on a pool of mock robots with user-configurable capabilities.

\subsection{A simple program}
\label{sec:hello}

\lstset{basicstyle=\small, showspaces=false, showstringspaces=false, numbers=left, numberstyle=\small, stepnumber=1, numbersep=-6pt, linewidth=33ex, language=Resh}

Here is a simple Resh program:
\begin{lstlisting}
  action say(string)
  task main() {
    say("I'm in the lobby!") @ "lobby"
  }
\end{lstlisting}
This program causes \emph{some} available robot, selected by the Resh runtime, to go to the lobby and there speak the given string.
The {\tt say} on line~3 is a {\bf robot action}; the {\tt @} clause specifies the location where the action is to be performed.
Locations in Resh are typically specified symbolically; the actual pose corresponding to a location is specified externally.

\lstset{numbers=none, linewidth=60ex}

In the Resh system, robots advertise the actions that they are capable of performing.
When an action is invoked in the code, that action will be assigned at runtime to a robot that advertises that capability.
This flexibility lets the same program be used in different situations with different types of robots.
If we do wish to fix a robot, we can use the Resh $\rightarrow$ construct:
\begin{lstlisting}
  say(...) @ "lobby" -> "robbie"
\end{lstlisting}
Now the action will be assigned to robbie.
If there is no robot named robbie in the pool, or if robbie is not currently able to drive to the lobby and speak, the action will not execute until robbie is in the pool and able to perform those actions.
Each robot must have a unique name, also advertised by the robot.

Robots (actually their agents; see \secref{agents}) can advertise whatever actions they wish, at whatever level of abstraction they wish.
An action advertisement is an assertion that the robot can perform the action autonomously (i.e., ``capability" = ``autonomously performable action").
Robots are responsible for performing their own actions --- Resh is not designed for tasks that need tightly controlled swarm behavior.

Because a robot's capabilities can change over time (sensors and actuators break or are added, environmental changes preclude or enable the performance of certain actions), advertisements can be made or retracted at any time.
Added actions are immediately eligible for use, even by currently running tasks.
Retracted actions that are currently in use can run to completion or fail, at the discretion of the robot.

Resh is strongly typed. 
The primitive 
types include {\bf robot}, a type {\bf loc} representing a geographic location, and the usual integer, boolean, and string types.
In the above program, line~1 declares that {\tt say} takes a single argument of type string.
Robots advertise actions along with a type signature.
Resh requires that when an action is invoked, it is assigned (by the runtime) to a robot that advertises
an action with the same name and type signature as the action declaration.
Advertisements may also contain a small amount of metadata such as the action's typical duration.

The Resh runtime has no semantic information about an action beyond its name, type signature, and metadata --- i.e., actions are \emph{uninterpreted}.
Semantic information about actions is carried in the documentation, not in Resh.  This use of uninterpreted actions corresponds exactly to the use of uninterpreted functions in typical programming languages.  
Two special actions are however interpreted by Resh: 
{\tt goto} and {\tt twist}, if advertised, must cause the robot to attempt to navigate to a specified position or perform a twist, respectively.
The runtime uses those actions to move the robots in their physical environment.

Resh is compositional: Tasks can call other tasks.
Further, there can be multiple top-level tasks executing on a given pool of robots at any given time.
The runtime is responsible for scheduling all the actions in all executing tasks onto the available robots.

Resh is deliberately not Turing-complete, to facilitate the automated synthesis of control strategies.
Resh is designed to be used in conjunction with an existing general-purpose language (GPL).
Such ``co-programming" is not a novel concept; a famous exemplar is Lex/Yacc, neither of which is Turing-complete.
The Resh system provides an API that enables GPL code to invoke, monitor, and interact with Resh tasks.

\subsection{A harder problem}
\label{sec:harder}

Now consider the following problem: \emph{Go pick up a package at location $A$ and deliver it to location $B$.}
We assume that workers at location $A$ load the robot,
and when they finish loading, they indicate that the robot can proceed to its destination.
Here is a complete Resh program for this scenario:

\lstset{numbers=left, linewidth=46ex}

\begin{lstlisting}
  action load(), dropoff()
  task main() {
    var r robot with !loaded
    waitevent pickup(A, B loc)
     => (load @ A ->r &
        waitprop r.loaded)
     => dropoff @ B ->r
  }
\end{lstlisting}

\lstset{numbers=none}

\vspace{-0.5ex}
\noindent
The program works as follows.
Line~1 declares the robot actions that we will use;
here {\tt load} and {\tt dropoff} make the robot do whatever it needs to do when loading and dropping off an item.
Line~3 declares a robot variable.
During execution, the runtime will choose an appropriate available robot and bind it to the variable.
This flexibility allows a Resh program to be run in multiple contexts.


The language semantics do not specify when the binding of variables to robots occurs.
In the current implementation the binding will not occur \emph{until the robot is actually needed}.
The choice of robot is determined by several heuristic factors, including proximity to the location(s) to which the program sends the robot.
The heuristics are not specified by the semantics and are thus free to change with improvements to the runtime.

A robot may have associated typed properties.
In line~3, the {\bf with} clause restricts the choice of robot to one with
a {\tt loaded} property whose current value is false.
Properties can be added, deleted, or modified at any time and are visible to all running programs.
The Resh runtime has no semantic information about properties beyond their name, type, and value.

Line~4 waits for the order to arrive, in the form of a Resh event.
The event, declared in the {\bf waitevent} construct, is named {\tt pickup} and has two arguments, which are the pickup and dropoff locations.

At the beginning of line~5 we see our first Resh temporal operator, $\Rightarrow$.
This operator represents simple temporal sequencing ---
after the expression on the left-hand side (LHS) finishes, the expression on the RHS will be started.
The language does not specify the start time of the RHS expression.
This flexibility helps the runtime schedule actions in a manner that best utilizes the available robots and best avoids action conflicts.

The parenthesized expression on lines 5 and~6 
contains our second temporal operator, \&.
This operator specifies that both operands must be performed, with no temporal ordering required between them.

Now consider the expression ``\lstinline{load @ A ->r}".
In \secref{hello} we saw an example of the $\rightarrow$ construct with a specific robot;
here we assign the {\tt load} action to whichever robot is bound to $r$.

Line~6 waits until the {\tt loaded} property on (the robot bound to) $r$ becomes true.
That is the indication that the workers have finished loading the robot.
Finally, line~7 drives the robot to the destination location and performs the {\tt dropoff} action there.
Because the {\tt load} and {\tt dropoff} actions are linked to the same robot variable,
it is guaranteed that the same robot that made the pickup also makes the dropoff.

This program requires that $r$ (1) has property {\tt loaded}, and (2) advertises {\tt load}, {\tt dropoff}, and {\tt goto} (the last because of the {\tt goto} implicit in the @ construct).
The runtime ensures that the robot chosen for $r$ satisfies (1) and (2).

\newcommand{\Va}{~$\vert$~}
\newcommand{\Vb}{~~$\vert$~~}
\newcommand{\Vc}{~~~$\vert$~~~}
\newcommand{\Opor}{\rule{.5pt}{1.5ex}}
\newcommand{\Opbp}{$!+$}
\newcommand{\Oppb}{$+!$}
\newcommand{\Opbpb}{$!+!$}
\newcommand{\Opba}{$!\&$}
\newcommand{\Opab}{$\&!$}
\newcommand{\Opbab}{$!\&!$}
\newcommand{\Opplusarrow}{$+\hspace{-5pt}\Rightarrow$}
\newcommand{\pc}[1]{\parbox{18pt}{\centering #1}}

%
\begin{figure}[htb!]\small
\begin{tabular}{lrll}
Action 		& $a$ 		& :=	& {\it id}~(~{\it arguments}~) \\
Waitevent	& $e$		& :=	& {\bf waitevent} {\it id}~(~{\it parameters}~) \\
Waitprop	& $p$		& :=	& {\bf waitprop} {\it propspec} \\
Pause		& $d$		& :=	& {\bf pause}~{\it duration} \\
Repeat		& $m$		& :=	& {\bf repeat} $t$ {\bf untilprop} {\it propspec} \\
Assign		& $r$		& :=	& [~~$\rightarrow$\Va $\leftrightarrow$~~]~~[~~{\it id}\Va {\it string}~~] \\
Location	& $c$		& :=	& @~[~~{\it id}\Va {\it string}~~] \\
Operator 	& {\it op} 	& := 	& \&\Vc $\Rightarrow$\Vc $+$\Vc \Opplusarrow\Vc \Opor \\
			&			& 		& \hspace*{-19ex}\Vc !\&\Vc \&!\Vc !\&!\Vc \Opbp\Vc \Oppb\Vc \Opbpb \\
Term 		& $t$ 		& := 	& $a$\Vc $e$\Vc $p$\Vc $d$\Vc $m$ \\
			&			& 		& \hspace*{-19ex}\Vc $t~~{\it op}~~t$\Vc (~$t$~)\Vc $t$~~[ $r$\Va $c$ ]
\end{tabular}
\caption{Resh expression grammar}
\label{fig:grammar}
\end{figure}

\begin{figure}[htb!]\small
$t_0$ \pc{\&} $t_1$ 			: $t_0$ and $t_1$ both execute, in no specified order\\
$t_0$ \pc{$\Rightarrow$} $t_1$ 	: $t_1$ starts sometime after $t_0$ finishes\\
$t_0$ \pc{$+$} $t_1$			: $t_0$ and $t_1$ start at nominally the same time\\
$t_0$ \pc{\Opplusarrow} $t_1$	: $t_1$ starts sometime after $t_0$ starts\\
$t_0$ \pc{\Opor}	$t_1$		: One of $t_0$ or $t_1$ executes\\
$t_0$ \pc{\Opba} $t_1$			: Same as \&, but $t_0$ is terminated (or not started) \\\hspace*{11.7ex}if $t_1$ finishes first\\
$t_0$ \pc{\Opab} $t_1$			: Same as \&, but $t_1$ is terminated (or not started) \\\hspace*{11.7ex}if $t_0$ finishes first\\
$t_0$ \pc{\Opbab} $t_1$			: Symmetric form of \Opba\ and \Opab\\
$t_0$ \pc{\Opbp} $t_1$, $t_0$~\Oppb~$t_1$, $t_0$~\Opbpb~$t_1$ : Same as $+$, but with term-\\\hspace*{11.7ex}ination as in the above forms of $\&$
\caption{Resh temporal operators}
\label{fig:ops}
\end{figure}

\subsection{Short-circuit operators}
\label{sec:short}

Sometimes we wish to prematurely terminate one of the operands of \& when the other completes.
In our example, suppose that when the robot is approaching location $A$,
the workers load it before it reaches $A$.
In this case we would like the robot to cease navigating to $A$ and proceed directly to $B$.
To implement this enhancement, we simply change the operator $\&$ on line~5 of the code in \secref{harder} to the short-circuit version \Opba.
Now if the RHS becomes true before the LHS finishes,
the driving and loading actions in the LHS will be prematurely terminated and the robot will proceed to its destination.

\subsection{Resh grammar}

Space precludes showing the entire Resh grammar;
\figref{grammar} shows a simplified form of the expression subset of the grammar.
Square brackets and long vertical bars are metasyntax. 
The {\bf pause} expression inserts a pause into the execution of the program;
{\bf repeat} repeats an expression until a property is satisfied.
The robot-assignment operator comes in two forms: the $\rightarrow$ form that we have seen and the exclusive form $\leftrightarrow$, which specifies that the robot must not be assigned any other action while it is executing the expression to which the $\leftrightarrow$ applies.
Robot assignments and locations may be applied to an expression as a whole, in which case they distribute to all actions in the expression.
There are 11~temporal operators, described in \figref{ops}.

\begin{figure}[htb!]
\centering
\includegraphics[width=30ex]{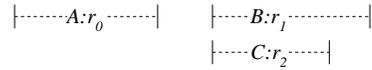}
\caption{Resh timing diagram}
\label{fig:timing}
\end{figure}

Consider the Resh expression ``\lstinline{A =>  (B + C)}".
\figref{timing} shows one possible timing diagram for these actions.
First $A$ is started and assigned to robot $r_0$.
Sometime after $A$ finishes, $B$ and $C$ are started at nominally the same time and assigned to other robots.
This illustration shows that actions have duration, the duration is unknown, and multiple actions may be started or may end at the same time.
Because the end time of an action cannot in general be predicted when the action is started,
Resh does not provide any temporal operators that place restrictions on the relative ordering of end times.

Resh timing diagrams can be encoded as sequences of words over an alphabet,
where each letter is a pair $\langle X,Y\rangle$ where $X$ represents a set of input action-termination events and $Y$ a set of output action-initiation events.
The Resh operator set is designed such that the set of all legal sequences for each Resh expression
is regular --- that is, it can be accepted by a finite automaton.
This property facilitates the synthesis of control strategies.

The Resh operator set was chosen for programming convenience, not minimality or orthogonality.
New operators may be added to the language for convenience, so long as they preserve regularity.
Space limits preclude a complete description of Resh semantics.

%
%
%

\section{THE RESH RUNTIME}
\label{sec:runtime}

\begin{figure}[htb!]
\centering
\includegraphics[width=35ex]{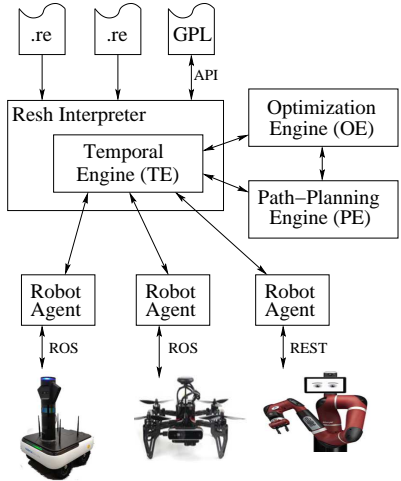}
\caption{The Resh system}
\label{fig:diagram}
\end{figure}

The Resh runtime is organized as shown in \figref{diagram}.
Each pool of robots has a single instance of the interpreter.
Any number of programs can be fed to the interpreter at any time; programs written in a general-purpose language interact with those in Resh via the API\@.
Three engines collaborate to run the given programs on the available robots in their current environment.
The interpreter controls the robots indirectly via agents,
which control the robots using robot-specific mechanisms;
\figref{diagram} shows two ROS-based robots and a third with a REST interface.
Unlabeled arrows in \figref{diagram} employ the Resh orchestration protocol, which is \emph{not} ROS-based.
As the focus in this paper is on the Resh language, we only briefly describe the roles of the engines and agents.


\subsection{Temporal Engine}
\label{sec:te}

The Temporal Engine (TE), contained in the Resh interpreter, is responsible for deciding which actions are currently eligible for execution and in which possible combinations.
For example, given the code ``\lstinline{A =>  (B + C)}",
if $A$ has finished executing, TE determines that $B$ and $C$ are now eligible for execution, but only if they are started at nominally the same time.

TE operates by watching for events that can possibly change the executability of actions.
These events include an action finishing executing (either successfully or unsuccessfully); the value of some robot property being changed; and a robot being added to, or deleted from, the available pool.

When one of these events occurs, TE formulates a mixed-Boolean-integer optimization problem \opt\ whose constraints encode the current state of execution of the programs, along with the robot capabilities and properties.
TE then hands \opt\ to OE to solve.
The solution specifies a (possibly empty) set of actions that should now start executing, and on which robots they should execute, and it may bind a subset of robot variables to specific robots.
When TE receives the solution, TE issues commands to start the specified actions on the specified robots.
Because TE re-optimizes on each event, the execution may not be globally optimal.

\subsection{Optimization Engine}
\label{sec:oe}

The Optimization Engine (OE) is responsible for solving \opt\@.
There may be a number of legal solutions.
For example, if the program simply specifies that some robot should go to location $L$ and there are two available fully-charged mobile robots, then assigning either robot to that action would be a valid solution.

OE is where the geometry of the situation is taken into account.
OE chooses among the possible solutions by considering geometry.
In our simple example, OE chooses the robot closer to $L$.
For more complex scenarios, OE takes into account the estimated travel times and battery life of all the robots. 

To determine travel times, OE formulates an instance of the multirobot-path-planning problem and gives it to PE\@.
For example, given the code ``\lstinline{A@L & B@M}"
(``do $A$ at location $L$ and $B$ at location $M$"), if OE has tentatively assigned $A$ and $B$ to robots $R$ and $S$, respectively, then OE formulates the problem ``send $R$ and $S$ from their current positions to locations $L$ and $M$, respectively."
Given solutions to these path problems, OE uses mixed-integer linear programming to choose from among its candidate solutions to \opt\ and sends the solution to TE\@.
The solution is guaranteed to satisfy the semantics of the executing code.


\subsection{Path-Planning Engine}

The Path-Planning Engine (PE) does high-level path planning for multiple autonomous robots.

Robots navigate by using, typically, a combination of a global and a local planner, with each robot using its own pair of planners.
These planners operate completely independently of the runtime, which has no visibility into, or control over, them.

PE performs path planning at a higher level, and for all available robots simultaneously.
PE attempts to find a set of sparse waypoints such that if the robots are sent to their ultimate destinations
via those waypoints, then the robots will not get too close to, or block, each other, and they will not travel too far from the ideal paths.
PE is also free to insert an arbitrary delay at each of the waypoints.

TE delegates to PE the job of assigning waypoints to robots (see \secref{msgflow}).
PE is also responsible for tracking the conformance of the actual paths taken by the robots with the intended paths.
The actual and intended paths can be different because
(1) The global and local planners solve different problems, with different inputs, using different algorithms, from those solved by PE;
and (2) There can occur transient obstructions or clearances in the environment that necessitate or enable different paths.
When the difference becomes too large, PE may abort some of the ongoing movements.
TE then formulates a new \opt\ based on the current state of the world and issues that problem as usual to OE\@.

\subsection{Robot agents}
\label{sec:agents}

Resh places no restrictions on what \emph{is} a robot.
Anything that advertises its capabilities to the runtime can be controlled by Resh programs.

To use a given robot with Resh, there must exist a Resh agent for that robot.
The agent is responsible for
(1) advertising the robot's name and capabilities to the runtime,
(2) publishing the robot status to the runtime, and
(3) translating between the Resh orchestration protocol and the robot-specific APIs.
For robots based on ROS, the agent interworks the Resh protocol with the protocols used on the ROS topics.
Resh can be used even with robots whose software stack is proprietary and closed.


The Resh system provides an API that makes it straightforward to implement agents.
As soon as the agent is up and the robot is added to the pool, programs may use the robot's capabilities, with zero changes required to the rest of the system.

\subsection{Message flow}
\label{sec:msgflow}

\begin{figure}[htb!]
\centering
\includegraphics[width=35ex]{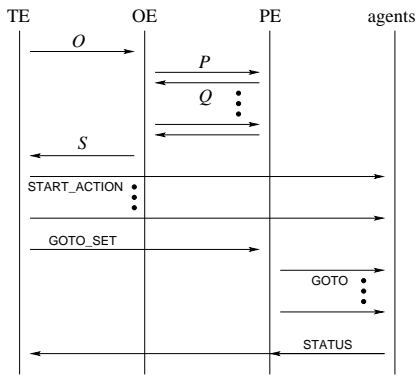}
\caption{Resh basic message flow}
\label{fig:msgflow}
\end{figure}

The Resh entities communicate with each other using a novel orchestration protocol.
The basic message flow is shown in \figref{msgflow}.
At opportune moments TE sends an optimization problem \opt\ to OE\@.
In the course of solving \opt, OE sends zero or more path problems $P$ to PE;
for each such problem PE returns a multipath solution $Q$.
OE eventually returns to TE a solution $S$ for the original problem.
$S$ contains zero or more action assignments.  
For each action that is not a {\tt goto},
TE sends a {\sc start\_action} message to the robot's agent.
For the {\tt goto}s, TE bundles them all up into a single {\sc goto\_set} message, which is sent to PE\@.
PE doles out those {\tt goto}s, possibly with delays added for purposes of collision avoidance.
The {\sc goto\_set} message also contains a reference to the multipath solution $Q$ that OE selected.

As the robots perform their actions and move in their environment, the agents send messages
indicating the status of those actions and the robot's current pose.
PE uses the status messages to track the conformance of the actual paths taken by the robots with the intended paths;
TE uses the status messages to update the temporal state of the executing Resh programs.

\section{IMPLEMENTATION}
\label{sec:impl}

Resh comprises 40k lines of code implemented in Go, Python, and other languages.
When a program is submitted to the system for execution, the parser
constructs the program's operator tree and the interpreter executes that tree.
A call from one task to another is handled by syntactically macro expanding the called task at each call site.
Recursion 
is not allowed.

The Optimization Engine uses the Python MIP package\cite{python-mip} for mixed-integer linear programming.
For testing purposes the system also provides a greedy, brute-force optimizer.

Agents for a number of different types of robots, both simulated and real, have been implemented, including:
a ROS-based Jackal UGV,
a Husarian ROSbot,
an Omron LD-90,
a Rethink Robotics Sawyer Arm,
a PX4-based Iris drone, 
and various research robots.
Agents for various devices that are not typically classified as robots have also been implemented,
and there is also a mock agent that controls no robot at all.
The remainder of the code is devoted to the Resh toolset, orchestration protocol, system APIs, and internal components.
The Resh API has a Go binding, with other bindings coming soon.

In our experiments in enterprise environments and robotic use cases around them, we have implemented task for a variety of scenarios, currently focused on surveillance and pickup-and-delivery.
The surveillance scenarios involve sending both specified and runtime-chosen robots of various types to locations in different geometrical and temporal patterns, and controlling the video feeds of those robots that are equipped with cameras.
The pickup-and-delivery scenarios also include interacting with the end user via speech generation and recognition.
Resh handles the action sequencing and invocation, while the actions themselves are performed by the robots.

\section{RELATED WORK}
\label{sec:related}

Resh is a domain-specific language for robotics.
But it is also a coordination language --- its main role being to coordinate the actions of a group of robots.
Coordination languages differ from traditional languages in that their focus is coordination and control rather than computation; they are thus not necessarily Turing-complete.
We relate Resh to three prominent streams of work on languages for coordination: process calculi such as Communicating Sequential Processes (CSP), behavior trees (BT) and finite-state machine (FSM) variants, and temporal logic.
Each has been a major source of inspiration in the development of Resh, although there are significant differences.

CSP was introduced in~\cite{DBLP:journals/cacm/Hoare78}.
Its core~\cite{DBLP:books/ph/Hoare85} is a calculus for the coordination of concurrent computations.
There is a crucial difference between the standard model of CSP and Resh: In CSP actions are atomic and (mathematically) instantaneous, whereas in Resh actions have a duration.
Conversion from atomic to non-atomic actions may introduce subtle errors~\cite{DBLP:conf/coordination/Kleine11}.
In Resh, on the other hand, nonatomicity is built into the semantics.
In~\cite{DBLP:conf/icra/TrckaS13}, a process algebra is used to specify tasks, and robots are assigned to (non-atomic) actions in a way that minimizes overall cost.
However, this approach assumes a fixed, known duration for each action and entirely abstracts the geometry of the physical space, which may limit practical application.

BTs are widely used in the game industry and have found application in robotics (cf.~\cite{iovino2020survey}).
In brief, a BT is a tree-shaped term built by combining basic actions with operators for sequencing, choice, fallback and concurrent execution.
This term is interpreted in rounds (``ticks") against the state of the world to determine which actions should be active.
Resh overlaps with BTs in several ways: Actions have duration and detectable failure modes, and the set of operators is similar.
A key difference is that a Resh program leaves open the assignment of robots to actions, giving the runtime considerable flexibility.

Linear temporal logic (LTL) is a well known notation for specifying concurrent behavior~\cite{DBLP:conf/focs/Pnueli77}.
It has found several applications to robotics, including in motion planning~\cite{DBLP:conf/icra/WolffTM13} and multi-robot coordination~\cite{DBLP:conf/icra/ElliottAHPT19,9197066}.
(These use either the GR(1) fragment or finite LTL, which have good synthesis algorithms.)
Unlike CSP and behavior trees, a temporal specification is not prescriptive: It merely defines a set of allowed execution sequences of actions, without fixing an execution strategy.
We have adopted this declarative approach in Resh. 
However, Resh differs from temporal logic in two key respects: one is that actions have duration (in temporal logic, propositions are atomic); the other is its support for sequential composition (not commonly included in temporal logic, cf.~\cite{DBLP:conf/lics/RosnerP86}).


The literature in robotic languages and multirobot task allocation are large; recent surveys are~\cite{10.1007/978-3-319-11900-7_17,khamis2015}.
Recent work related to Resh is as follows.
Buzz \cite{7759558} is designed for heterogeneous teams but has assumptions that apply only to swarms, which are out of scope for Resh.
NVL \cite{marques2015} and Dolphin \cite{8594059} are Turing-complete languages that contain features analogous to Resh, but the systems require user code to do its own scheduling of robots to actions.
The language of \cite{8462885} is XML-based and targets a different class of problems 
from Resh.
XRobots \cite{6225145} is based on hierarchical FSMs.
vTSL \cite{8593559} is based on the verification language Promela and does not (as of the cited publication) provide a way to run the programs on robots.
TeCoLa \cite{koutsoubelias2016} bears strong architectural similarity to the Resh system; it also provides an elegant subteam mechanism.
However, TeCoLa takes the API approach instead of a language.
It also requires user code to do its own scheduling, and it does not support multiple simultaneous tasks.
Another recent API-based system is~\cite{yi2020}.
The systems \cite{7759698,9197527} do not provide languages, but use algorithms that combine task and motion planning in a manner similar to the Resh runtime.
In \cite{9197283} the authors discuss the dynamic nature of robot capabilities, but the focus is the optimization algorithm rather than languages or APIs. Planning systems based on notations such as PDDL~\cite{pddl-1998} require an accurate world-representation and a precise semantics for each action, which may be difficult to come by, thus Resh leaves most actions uninterpreted.
Finally, the VDA 5050 standard \cite{vda5050-1.1} defines a robot-orchestration protocol that has a number of similarities with the Resh protocol.
 
\section{CONCLUSIONS}
\label{sec:concl}

Resh is a new typesafe programming language for orchestrating multirobot systems.
The Resh runtime combines a number of common algorithms to support the execution of Resh programs and thus removes that burden from a programmer.
We use Resh to program a complex and frequently changing multirobot system; in our experience, it has considerably simplified the programming task and made it more enjoyable.

Resh is a work in progress and as such has a number of deficiencies.
The current runtime can assign actions to robots in a manner that is suboptimal in the long term.
Work on implementing deeper lookahead strategies using program synthesis (cf.~\cite{DBLP:conf/popl/PnueliR89}) is ongoing.
Finally, Resh's notion of error handling is minimalistic:
If an action fails, the entire containing task is aborted.
We plan to extend Resh to support more sophisticated error handling.

\section*{ACKNOWLEDGMENT}
\label{sec:ack}
The authors thank Al Aho, Rodney Brooks, Fangzhe Chang, Tom Van Cutsem, Manuel Deneu, Katherine Guo, Ilija Had\v{z}i\'{c}, Dong Liu, and Hans Woithe for their help in the design and implementation of Resh and for their feedback on this paper.

\addtolength{\textheight}{-9.2cm}

\bibliographystyle{plain}
\bibliography{robotics.bib}

\end{document}